# Image Steganography, a New Approach for Transferring Security Information

H.B. Bahar[†] and Ali Aboutalebi[††]
*Tabriz University*
*Faculty of Electrical and Computer Engineering*

*July 2008*

[†] Email: h.b.bahar@tabrizu.ac.ir

[††] Email: abootalebi@gmail.com

*Image Steganography, a New Approach for Transferring Security Information*

**H.B. Bahar and Ali Aboutalebi**
*Tabriz University*
*Faculty of Electrical and Computer Engineering*

# ABSTRACT

Steganography is the art of hiding the fact that communication is taking place, by hiding information in other information. Many different carrier file formats can be used, but digital images are the most popular because of their frequency on the Internet. For hiding secret information in images, there exists a large variety of steganographic techniques some are more complex than others and all of them have respective strong and weak points. Different applications have different requirements of the steganography technique used. For example, some applications may require absolute invisibility of the secret information, while others require a larger secret message to be hidden. This paper intends to give an overview of image steganography, its uses and techniques. It also attempts to identify the requirements of a good steganographic algorithm and briefly reflects on which steganographic techniques are more suitable for which applications [1].

# 1. Introduction

The techniques of Steganography are classified into linguistic steganography and technical steganography [2]. The former consists of linguistic or language forms of hidden writing. The later, such as invisible ink, try to hide messages physically. One disadvantage of linguistic steganography is that users must equip themselves to have a good knowledge of linguistry. In recent years, everything is trending toward digitalization. And with the development of the Internet technologies, digital media can be transmitted conveniently over the network. Therefore, messages can be secretly carried by digital media by using the steganography techniques, and then be transmitted through the Internet rapidly.
The steganographic terminologies used in this paper agreed with those in [3]. The goal of steganography is covert communication. So, a fundamental requirement of a steganographic system is that the hidden message carried by stego-media should not be sensible to human beings. The most general steganographic model presented by G.J. Simmons is the prisoners' problem [4]. In this problem, two persons in the jail plan to make an escape together. All communications between them are monitored by the warden. So they must hide the messages concerning escape plan in another innocuous- looking media. An assumption in this model is that both the sender and receiver must have shared some secret information before imprisonment. So the prisoners' problem is classified into secret key steganography. Pure steganography means that there is none prior information shared by two communication parties. If the public key of the receiver is known to the sender, the steganographic protocol is called public key steganography [5].
The warden may be passive, that is, he only observes the passing messages. If the warden detects the occurrence of covert communication, the prisoners will be frustrated in their attempt to escape and will be thrown into solitary confinement. Steganalysis is the art of discovering the existence of hidden information. In [6], Johnson and Jajodia identify some characteristics of stego-images that are created by specific image steganographic systems.
To remove all possible covert messages, an active warden may be allowed to slightly modify the data being sent between prisoners. An example of mild modification per-formed by the active warden is to replace the words with some close synonyms in the mail documents. If the carrier of secret messages is an image, any low-pass filters can be utilized for obviating covert communication. It is worthy to note that the primary goal of an active warden is to avoid covert communication taking place. On the other hand, in the real world a passive warden or monitor makes an attempt to find unknown criminals from their communication to a known criminal. Opposite to the goal of steganalysis, the requirements of a steganographic system include not only imperceptibility but also un- detectability by any steganalysis tool. When examined by an active warden, the hidden message should be robust against any possible modification. There are some steganographic protocols in the presence of a passive warden or an active one are described in [7-8].

The most common and simplest image embedding method is the least significant bit (LSB) insertion. The LSB insertion embeds message in the least significant bit of some selected pixels. In this scheme, the embedding capacity can be increased by using two or more least significant bits. At the same time, not only the risk of making the embedded message statistically detectable increases but also the image fidelity degrades. Hence a variable-sized LSB embedding scheme is presented in [9-10], in which the number of LSBs used for message embedding/extracting depends on the local characteristics of the pixel.

The implementation of LSB scheme must adapt to diverse media file formats. There are many image steganographic techniques that strongly correlated with the format of the cover-image. S-Tools [11], Stego [12], and Fridrich's method [13] are designed for palette-based images, and Jpeg-Jstego [14] is used for JPEG compressed images. Consequently, format conversion of these stego-images will destroy the whole hidden message. And slightly modify the palette in the pallette-based image or recompress the JPEG images using different quality, i.e., change the quantization table; will destroy the whole hidden message too.

The advantages of LSB-based method are easy to implement and high message pay-load. Unfortunately, the hidden message is vulnerable to even a slight modification from a active warden. Marvel et al. [15] present an image steganographic method, entitled spread spectrum image steganography (SSIS), that hides and recovers the message within digital imagery. The SSIS incorporated the use of error-control codes to correct the large number of bit errors. The performance of SSIS has been demonstrated by adding low levels white Gaussian noise and by applying low levels of JPEG compression. With Q-factor of 80, the resulting embedded signal BER is 0.3001. Consequently, the (2040, 32) binary expansion of Reed-Solomon code must be used for error-free message recovery. In the next Section, we will describe our proposed blind secret-key image steganographic model. Some experimental results and performance analysis will be presented in Section 3 to show that the proposed model is practicable. Finally in Section 4, conclusions and future works will be presented [15].

## 2. Image Steganography and bitmap pictures

Using bitmap pictures for hiding secret information is one of most popular choices for Steganography. Many types of software built for this purpose, some of these softwares use password protection to encrypting information on picture. To use these softwares you must have a "BMP" format of a picture to use it, but using other type of pictures like "JPEG", "GIF" or any other types is rather or never used, because of algorithm of "BMP" pictures for Steganography is simple. Also we know that in the web most popular of image types are "JPEG" and other types not "BMP", so we should have a solution for this problem. In next section of this paper we have brief case of using "BMP" format algorithm and after that we offer a solution for solving discussed problem.

## 3. Bitmap Steganography

Bitmap type is the simplest type of picture because that it doesn't have any technology for decreasing file size. Structure of theses files is that a bitmap image created from pixels that any pixel created from three colors (Red, Green & Blue briefly said RGB) each color of a pixel is one byte information that shows the density of that color. Merging these three colors makes every color that we see in these pictures. We know that every byte in computer science is created from 8 bit that first bit is Most-Significant-Bit (MSB) and last bit is Least-Significant-Bit (LSB), the idea of using Steganography science is in this place; we use LSB bits for writing our security information inside BMP pictures. So if we just use last layer (8st layer) of information, we should change the last bits of pixels, in other hands we have 3 bits in each pixel so we have **3*height*width** bits memory to write our information. But before writing our data we must write name of data (file), size of name of data & size of data. We can do this by assigning some first bits of memory (8st layer).

```
(00101101    00011101    11011100)
(10100110    11000101    00001100)
(11010010    10101100    01100011)
```

*Figure1. Using each 3 pixel of picture to save a byte of data*

$$Storage\ size\ in\ each\ layer = \frac{3 * width * height}{8}\ (bytes)$$

## 4. Using other type of pictures for Steganography

Now we come back to our problem that using other types of picture. To solve this I discuss a technology about software-programming called Microsoft .NET.

Microsoft .NET framework prepares a huge amount of tools and options for programmers that they simples programming. One of .NET tools for pictures and images is auto-converting most types of pictures to BMP format. I used this tool in a software called "AnyFile2Image" [16] that is written in C#.NET language and you can use this software to hide your information in any type of pictures without any converting its format to BMP (software converts inside it).

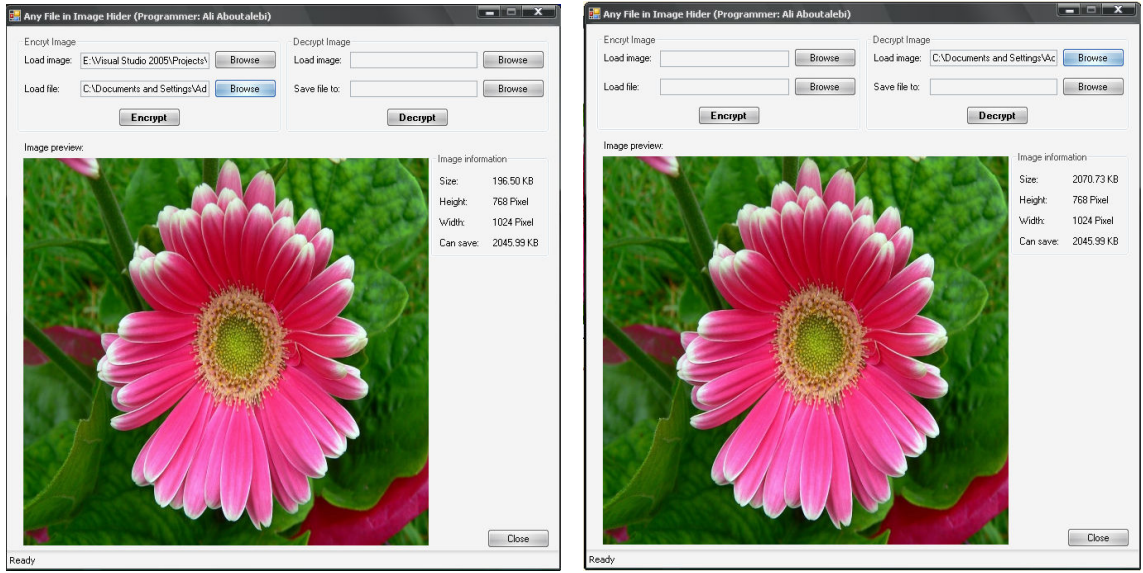

*Figure 2. Encryption (left) and decryption (right) steps in AnyFile2Image software*

## 5. Encrypting and decrypting algorithm

The algorithm used for Encryption and Decryption in this application provides using several layers lieu of using only LSB layer of image. Writing data starts from last layer (8st or LSB layer) because significant of this layer is least and every upper layer has doubled significant from its down layer. So every step we go to upper layer image quality decreases and image retouching transpires.

Before encrypting file inside image we must save name and size of file in a definite place of image. We could save file name before file information in LSB layer and save file size and file name size in most right-down pixels of image. Writing these informations is needed to retrieve file from encrypted image in decryption state.

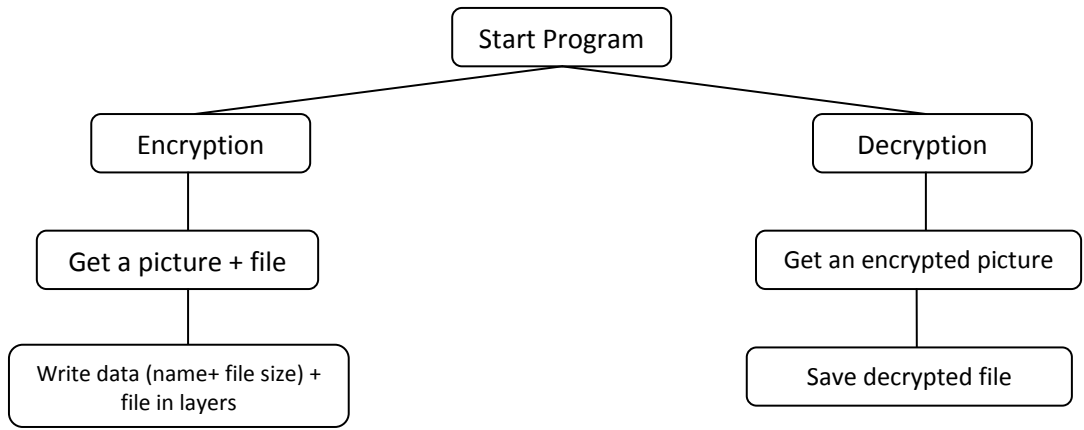

*Figure 3. Basic structure of application*

# 6. Conclusion

In this paper we discussed about a solution of Steganography science to how to use any type of image formats to hiding any type of files inside their. "AnyFile2Image" [16] software provides this purpose for users. For improvement of this base application we can add password protection options to have a secure information transmission.

There are several softwares like "AnyFile2Image" that takes good choices for user to hide his/her information in picture [17], but masterwork of "AnyFile2Image" is in supporting any type of pictures without need to convert to bitmap, and lower limitation on file size to hide, because of using maximum memory space in picture to hide the file.

Since ancient times, man has found a desire in the ability to communicate covertly. The recent explosion of research in watermarking to protect intellectual property is evidence that steganography is not just limited to military or espionage applications. Steganography, like cryptography, will play an increasing role in the future of secure communication in the "digital world" [18].

# References

[1] T.Morkel , J.H.P. Eloff , M.S.Olivier - "An overview of image Steganography" - Information and Computer Security Architecture (ICSA) Research Group Department of Computer Science University of Pretoria, 0002, Pretoria, South Africa.

[2] F. L. Bauer, "Decrypted Secrets - Methods and Maxims of Cryptology," Berlin, Heidelberg, Germany, Springer-Verlag (1997).

[3] Birgit Pfitzmannn, "Information Hiding Terminology", in Proceedings of FirstWork- shop of Information Hiding, Cambridge, U.K. May 30 - June 1, 1996. Lecture Notes in Computer Science, Vol.1174, pp 347-350. Springer-Verlag (1996).

[4] Gustavus J. Simmons, "The Prisoners' Problem and the Subliminal Channel", in Proceedings of CRYPTO '83, pp 51-67. Plenum Press (1984).

[5] Stefan Katzenbeisser and Fabien A. P. Petitcolas, "Information Hiding Techniques for Steganography and Digital Watermarking," Artech House (2000).

[6] Neil F. Johnson and Sushil Jajodia, "Steganalysis of Images Created using Current Steganography Software," in Proceedings of 2nd International Workshop on Information Hiding, April 1998, , Portland, Oregon, USA. pp. 273 - 289.

[7] Ross J. Anderson and Fabien A.P. Petitcolas, "On the limits of steganography," IEEE Journal on Selected Areas in Communications (J-SAC), Special Issue on Copyright & Privacy Protection, vol. 16 no. 4, pp 474-481, May 1998.

[8] Scott Craver, "On Public-key Steganography in the Presence of an Active Warden," in Proceedings of 2nd International Workshop on Information Hiding, April 1998, Portland, Oregon, USA. pp. 355 - 368.

[9] Yeuan-Kuen Lee and Ling-Hwei Chen, "An Adaptive Image Steganographic Model Based on Minimum-Error LSB Replacement," in Proceedings of the Ninth National Conference on Information Security, pp. 8-15. Taichung, Taiwan, May 14-15, 1999.

[10] Yeuan-Kuen Lee and Ling-Hwei Chen, "A High Capacity Image Steganographic Model," accepted by IEE Proceedings Vision, Image and Signal Processing. (2000).

[11] A. Brown, "S-Tools", Shareware ftp://idea.sec.dsi.unimi.it/pub/security/crypt/code/s-tools4.zip (Version 4).

[12] Romana Machado, "Stego", Shareware, http://www.stego.com.

[13] Jiri Fridrich, "A New Steganographic Method for Palette-Based Images", in Proceedings of the IS&T PICS Conference, Savannah, Georgia, April 25-28, 1999, pp.285-289.

[14] D. Upham, Jpeg-Jstego, Modification of the Independent JPEG Group. JPEG software (release 4) for 1-bit steganography in JFIF output file. ftp://ftp.funet.fi/pub/crypt/steganography.

[15]    Yeuan-Kuen Lee and Ling-Hwei Chen - "A Secure Robust Image Steganographic Model" - Department of Computer and Information Science. National Chiao Tung University, Hsinchu 30050, Taiwan, R.O.C.

[16]    Source of Software URL:  http://sourceforge.net/projects/AnyFile2Image/

[17]    Image Steganography softwares like "Hide in Picture" (see: http://sourceforge.net/projects/hide-in-picture/).

[18]    Eugene T. Lin and Edward J. Delp, "A Review of Data Hiding in Digital Images", Video and Image Processing Laboratory (VIPER), School of Electrical and Computer Engineering, Purdue University West Lafayette, Indiana.